\documentclass[preprint2]{aastex}
\usepackage{natbib}

\newcommand{\gtsim}{\mbox
{{\raisebox{-0.4ex}{$\stackrel{>}{{\scriptstyle\sim}}$}}}}
\newcommand{\ltsim}{\mbox
{{\raisebox{-0.4ex}{$\stackrel{<}{{\scriptstyle\sim}}$}}}}

\slugcomment{Post-ref version}

\shorttitle{Blundell \& Rawlings}
\shortauthors{The spectra of lobes}

\begin{document}

\title{The spectra and energies of classical double radio lobes}

\author{Katherine M.\ Blundell and Steve Rawlings}
\affil{University of Oxford, Astrophysics, Keble Road, Oxford,
OX1 3RH, UK}

\begin{abstract} We compare two temporal properties of classical double
radio sources: i) radiative lifetimes of synchrotron-emitting particles
and ii) dynamical source ages.  We discuss how these can be quite
discrepant from one another, rendering use of the traditional spectral
ageing method inappropriate: we contend that spectral ages give meaningful
estimates of dynamical ages only when these ages are $\ll 10^7$\,years.
In juxtaposing the fleeting radiative lifetimes with source ages which are
significantly longer, a refinement of the paradigm for radio source
evolution is required.  We move beyond the traditional bulk backflow
picture and consider alternative means of the transport of high Lorentz
factor ($\gamma$) particles, which are particularly relevant within the
lobes of low luminosity classical double radio sources.  The changing
spectra along lobes are explained, not predominantly by synchrotron ageing
but, by gentle gradients in a magnetic field mediated by a low-$\gamma$
matrix which illuminates an energy-distribution of particles, $N(\gamma)$,
controlled largely by classical synchrotron loss in the high magnetic
field of the hotspot.  A model of magnetic field whose strength lowers
with increasing distance from the hotspot, and in so doing becomes
increasingly different from the equipartition value in the head of the
lobe, is substantiated by constraints from different types of
inverse-Compton scattered X-rays.  The energy in the particles is an order
of magnitude higher than that inferred from the minimum-energy estimate,
implying that the jet-power is of the same order as the accretion
luminosity produced by the quasar central engine.  This refined paradigm
points to a resolution of the findings of Rudnick et al (1994) and
Katz-Stone \& Rudnick (1994) that both the Jaffe-Perola and
Kardashev-Pacholczyk model spectra are invariably poor descriptions of the
curved spectral shape of lobe emission, and indeed that for Cygnus A all
regions of the lobes are characterised by a `universal spectrum'.
\end{abstract}


\section{Introduction: lobe loss mechanisms}
\label{sec:intro}
That the radio radiation from classical double radio sources
arises from synchrotron emission is beyond doubt: the spectra of
many emitting regions of these sources are predominantly
power-law and often characterised by a high fractional
polarisation.  Synchrotron cooling is by no means the only
significant loss mechanism to be manifested on the spectra of
these sources however.  A recent study has demonstrated the
importance of expansion losses on evolving radio sources, which
we briefly review.  

\citet*[figure 8]{Blu99a} found that the strongest
correlation\footnote{The strongest intrinsic correlation that is; a very
strong artificial correlation is found between luminosity and redshift
because of the Malmquist bias.} between any two properties of
low-frequency selected radio sources in complete samples is that sources
with longer linear sizes $D$ have steeper spectra $\alpha$ when these are
evaluated at rest-frame 151\,MHz.  This is known as the $D$--$\alpha$
correlation.  This correlation is independent of any separate correlation
between other source properties: the Spearman rank correlation coefficient
at constant luminosity $P$ and constant redshift $z$ is $r_{D\alpha \vert
Pz} = 0.37$, with any dependence on the assumed cosmological model only
arising in the significance of this effect (which varies between $5\sigma$
-- $6\sigma$ for the models considered).  Moreover, simulations of model
sources discussed in that paper show that throughout the lifetime of an
individual source, its low-frequency spectrum steepens.  These
simulations, when combined with the sampling-functions due to the survey
flux-limit and light-cone interception, regenerate the $D$--$\alpha$
correlation.

\citet{Blu99a} attributed the steepening of low-frequency spectra with age
to adiabatic expansion losses.  Adiabatic expansion losses occur when an
emitting blob of plasma expands, with a consequent decrease in the
magnetic field strength and in the energies of the particles themselves
\citep{Sch68}.  When one considers a fixed observing frequency, a lower
magnetic field means that higher Lorentz factor particles are required to
radiate at the chosen frequency.  The relationship between the Lorentz
factor $\gamma$ of particles giving rise to emission at some specific
frequency $\nu$, and the ambient magnetic field strength $B$ is given by:
\begin{equation}
\label{eq:lorentz}
\gamma = \biggl(\displaystyle\frac{m_{\rm e}}{eB} 2\pi \nu \biggr)^{
          \frac{1}{2}},
\end{equation}
where $e$ is the charge on an electron and $m_{\rm e}$ is its
rest-mass.

Because it is essential to factor in not just synchrotron losses (and, at
high redshift, inverse Compton losses off the cosmic microwave background
[CMB]) but also adiabatic expansion losses to analyses of the spectra of
radio lobes, we consider in this paper how these arise and what changes
must be made to existing views of radio source evolution.  The outline of
this paper is as follows: in \S\ref{sec:specage} we describe briefly
studies to date which have inferred spectral ages of classical double
radio sources, in \S\ref{sec:radlife} we specify and quantify the
radiative lifetimes of synchrotron particles in the lobes, in
\S\ref{sec:sourcelife} we discuss what is meant by the lifetime of a radio
source, and then in \S\ref{sec:sourceage} we compare the dynamical ages of
sources we observe with their spectral ages and with the relevant
radiative lifetimes.  In \S\ref{sec:taildock} we re-visit the problem,
first identified by \citet{Jen76}, that longer and less-powerful radio
sources invariably have lobe emission (even at GHz frequencies) which
often persists back to the core, while shorter and more-powerful sources
do not have lobe emission which extends back to the core.  We discuss
seven mechanisms in \S\ref{sec:mix} by which the findings of the two
previous sections might be reconciled.  One of these mechanisms has been
almost completely neglected in the past and is worthy of further serious
consideration.  We discuss possible information which might be gleaned
from the rarity of `dead' radio galaxies, discussed in \S\ref{sec:relic}.
In \S\ref{sec:refine} we discuss a refined picture of radio source growth.

Our refined model enables us to identify a contribution --- more important
than just synchrotron cooling --- to spectral steepening which is observed
to increase along radio lobes with increasing distance from the hotspot;
this we discuss in \S\ref{sec:specgrad}.  In \S\ref{sec:newbit} we
investigate independent support for the predictions of this model, from
detections of inverse-Compton X-rays associated with radio sources and
consider the implications for the overall energy budget.

We examine the results of \citet{Rud94} and \citet{Kat94} in
\S\ref{sec:katz} in the light of the model which we discuss in
\S\ref{sec:refine}.  We conclude in \S\ref{sec:conc}.

\section{Spectral ageing}
\label{sec:specage}
If a magnetized blob of plasma, with an initial power-law spectrum
$N(\gamma)$ to infinite energy, were subject only to synchrotron cooling
via the uniform magnetic field strength ($B$) in which it is deemed to be
immersed, then a cutoff or `break' in that power-law spectrum would appear
(Kardashev 1962, Pacholcyzk 1970, Jaffe \& Perola 1973).  The frequency at
which this break occurs ($\nu_{\rm B}$) would move to lower frequencies as
$\nu_{\rm B} \propto B^{-3}t^{-2}$ where $t$ is the time elapsed since the
blob of plasma had its particles accelerated to their initial power-law
energy distribution.  Fits to the curved spectra observed in small
transverse elements of radio lobes have led changes in spectral curvature
to be identified with these break frequencies.  These estimates of the
break frequencies, together with an estimate of the magnetic field
strength have led to estimates of the `spectral age' [or time elapsed
since acceleration] being made across successive elements in the lobes of
radio sources \citep[e.g.\ ][]{Win80,Mye85,Ale87b,Liu92}.  These workers
have interpreted the spectral ages of the regions of lobe nearest the core
as being closely related to the source age.

However, if other loss mechanisms besides synchrotron cooling, such as
adiabatic expansion losses as discussed in \S\ref{sec:intro}, are also
taking place then the observed break frequency in the spectrum will have
been reached more rapidly than the above simple picture will allow, as
long as the magnetic field strength has only tended to decrease with time.
Unless the true value of the magnetic field is vastly less than that
estimated (see discussion in \S\ref{sec:sub_equip}), this would render
simple spectral ages to be {\em over-estimates} of the time elapsed since
acceleration \citep[see e.g.\, ][]{Ale87a}.  It is likely that a given blob
of plasma will have spent some time in a higher magnetic field when it was
near the hotspot than subsequently for example, and knowledge of the
`history' of the magnetic field evolution is required to begin to
re-interpret measured spectral ages.  A modification of the formula for
break frequency when both synchrotron and adiabatic expansion losses take
place may be found in \citet{Blu94b}.

Many have aired concerns about the spectral ageing method, e.g.\
\citet*{Jon99} have pointed out that any {\em mixing} of particles within
the lobes will contaminate estimates of spectral ages, while \citet{Tri93}
has demonstrated that subtleties in spectral evolution are more
appropriately attributed to the magnetic field configuration than to
differences in the evolution of the pitch-angle distribution, while others
have discussed the difficulty of estimating the magnetic field strengths
or identifying the magnetic field history for a given blob of plasma
\citep[e.g.\ ][]{Sia90,Wii90,Rud94,Rud99} and we do not repeat them here.
We remark, however, that unless the spectral ages to be measured are much
shorter than the radiative lifetimes of the particles which constitute the
evolving spectrum, spectral ageing is a flawed method.

\section{Radiative lifetimes}
\label{sec:radlife}

We now consider for how long a synchrotron particle can
contribute to emission observed from a radio lobe.

A very naive estimate of the length of time for which a
synchrotron-emitting electron will radiate, under the conditions
thought to be found in a lobe of a classical double radio source,
may be obtained by dividing the total kinetic energy $E$, of
one electron by its synchrotron power $P$, as follows:
\begin{equation}
\label{eq:syncloss}
{\displaystyle \frac{E}{P} = \frac{(\gamma - 1) m_{\rm e} c^2}
                             {2 \sigma_{\rm T} c \gamma^2
\biggl(\displaystyle{\frac{v}{c}}\biggr)^2 U_{\rm mag} \sin^2\theta}}.
\end{equation}
Here $\gamma$ is the Lorentz factor of the electron, $c$ is the speed of
light, $\sigma_{\rm T}$ is the Thompson scattering cross-section for
electrons, $v$ is the velocity of the electron, $U_{\rm mag}$ is the
magnetic energy density and $\theta$ is the pitch-angle of the electron's
motion with respect to the magnetic field lines.  We set $v/c$ and
$\sin^2\theta$ to 1.  Equation~\ref{eq:lorentz} tells us that to radiate
at rest-frame 5\,GHz in a magnetic field strength of 1\,nT the required
Lorentz factor of the electron is $\gamma \sim 10^4$.

\begin{figure}[!h]
\begin{picture}(50,395)(0,0)
\put(-77,-25){\includegraphics{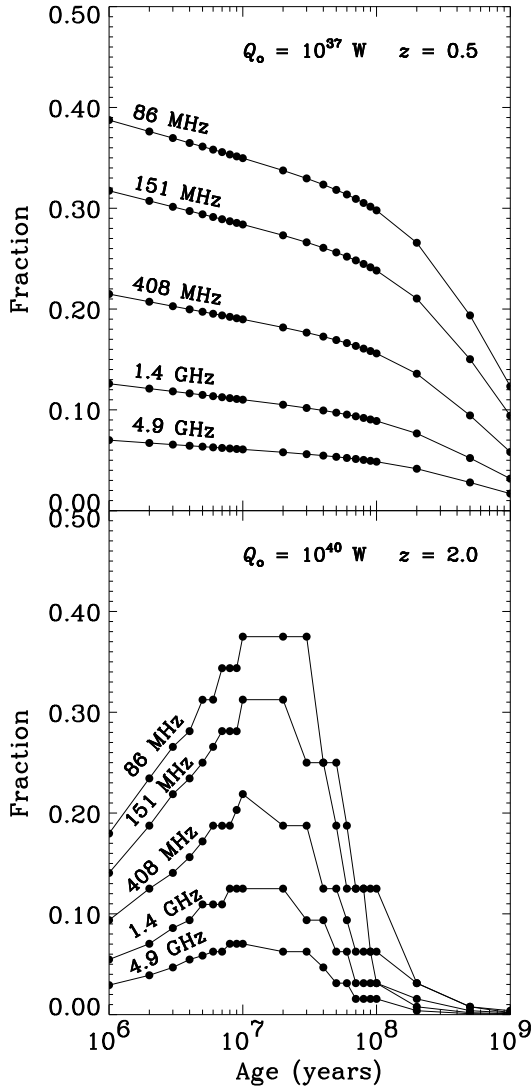}}
\end{picture}
{\caption[junk]{\label{fig:tmin} These plots show the {\em fraction of the
age} of the radio source for which the particles radiating at a certain
frequency have been in the radio lobe, against the age of the
radio-source.  The lower plot is for a simulated radio source with each
jet carrying a bulk kinetic jet-power $Q_{\rm o} = 10^{40}\,{\rm W}$ and
the upper plot is for $Q_{\rm o} = 10^{37}\,{\rm W}$.  The assumed
environment for these simulated radio sources has a number density at
radius 100\,kpc of $n_{100} = 2 \times 10^3\,{\rm m^{-3}}$ and a
spherically symmetric profile outside a core radius of $r = 10$\,kpc whose
density falls off as $r^{-1.5}$.  The normalisation factor $c_1$ for the
source growth in the model of \citet{Fal91} is 3.5 (Kaiser et al.\ 1997).
}}
\end{figure}

The radiative lifetime of such an electron would be just under $10^7$
years according to Equation~\ref{eq:syncloss} with $\gamma = 10^4$.
However, this naive sum overlooks two important factors: i) clearly a
particle with $\gamma \sim 10^4$ cannot radiate for $10^7$ years at 5\,GHz
in the assumed constant magnetic field.  Rather, this duration of time
indicates how long it would take a particle cooled by synchrotron losses
to change from having $\gamma \sim 10^4$ to having $\gamma \sim 1$ if its
cooling rate were maintained at the level expressed in
equation~\ref{eq:syncloss}, with $\gamma = 10^{4}$.  ii) The lobe of a
radio source is an evolving entity and the adiabatic expansion losses
(especially the decreasing magnetic field) discussed earlier mean that
particles with continuously higher $\gamma$ are required to produce the
emission at the same chosen frequency as the source ages.  For example, in
the scenario above, a $\gamma \sim 10^4$ particle will emit most of its
radiation at 5\,GHz if the magnetic field strength is 1\,nT.  If the field
lowered adiabatically to 0.1\,nT then particles whose Lorentz factors are
$3 \times 10^4$ are required to produce the emission at the same
`observing frequency' of 5 GHz.  In a radio lobe, changes in the magnetic
field go hand-in-hand with changes in the energies of the particles
themselves since it is the expansion of the lobe which governs both.  Thus
the particle now radiating in the lower $B$-field of 0.1\,nT at 5\,GHz
with $\gamma \sim 3 \times 10^4$ would, prior to the decrease in magnetic
field strength, have had $\gamma \sim 10^5$.

\citet{Mat90} were the first to consider the losses suffered by particles
of a given Lorentz factor ($\gamma$) due to synchrotron cooling and
adiabatic expansion losses as a function of time and to quantify their
influence on ${\rm d}\gamma/{\rm d}t$.  \citet*{Kai97b} have modified
their relation to include the influence of inverse-Compton losses off the
CMB (which follow the behaviour of synchrotron losses) and this may be
used to relate the Lorentz factor of a synchrotron particle contributing
to the emission at some given frequency, the time it was injected into the
lobe and its Lorentz factor at the time it was injected.  The relation can
be used to define a maximum time which can elapse between an initial time
when a particle is injected with Lorentz factor $\gamma$ and the time of
observation, if the particle is to emit synchrotron radiation at the {\em
specified observing frequency}.  This time of injection is called $t_{\rm
min}$; prior to this time, even if particles of extremely high Lorentz
factor are injected into the lobe, their enhanced energy losses will be so
catastrophic that their Lorentz factors will be too low at the time of
observation\footnote{We use the term `time of observation' quite liberally
here to mean `when the source intercepts our light-cone' or equivalently
`when the light we ultimately observe leaves the source'.}  ($t_{\rm
obs}$) to contribute to radiation at the chosen frequency, in a given
$B$-field.

Thus for the ensemble of particles contributing to the radiation
at a given emitted frequency at $t_{\rm obs}$, those which had
the largest Lorentz factors at injection are injected at $t_{\rm
min}$ and those with the lowest are those actually injected at
$t_{\rm obs}$.  There is no trivial identity which connects
$t_{\rm min}$ with the age of the source, as Fig.~\ref{fig:tmin}
shows.

The clear implication from this figure is that a few $10^7$ years is a
very definite upper limit to the maximum radiative lifetime of a
synchrotron emitting particle, if the magnetic field experienced by the
particles in the lobe is as we describe in \S\ref{sec:equip} (namely
gently decreasing because of the expansion of the lobe, and of order a few
nT).  Particles responsible for radiation at lower frequencies have a
greater difference between $t_{\rm min}$ and $t_{\rm obs}$ as seen in
Fig.~\ref{fig:tmin} and discussed in \citet{Blu99a}.

Thus reconciling inferred spectral ages (discussed in \S\ref{sec:specage})
(particularly if synchrotron cooling is the only loss mechanism accounted
for) with source ages (see \S\ref{sec:sourceage}) necessarily requires a
re-think, since any given set of particles only contributes to the
radiation for a small fraction of the age of a classical double.  We
discuss such a re-think in \S\ref{sec:mix}.

\section{Source lifetimes} 
\label{sec:sourcelife}
Given current terminology in the literature we first clarify our useage of
the terms `age' and `lifetime' in the context of observed radio sources.
Knowledge of their {\em lifetimes} --- the maximum duration of time for
which sustained accretion and jet production is to be maintained --- is
relevant for constraining models of jet-producing central engines.
Current modelling of the maximum duration of time for which sufficient
jet-power may be extracted from the black-hole suggests limits of $10^8$
-- $10^9$ years \citep{Mei99} which depend on both the mass and the spin
of the black hole.  Knowledge of the {\em ages} of the observed objects
under study is crucial if one is to correctly interpret details of the
environmental context in which the radio source resides.

\section{Source ages} 
\label{sec:sourceage}
The simplest constraints on source ages come from making use of the
measured projected physical size of a radio galaxy in a particular
cosmological model and estimating the speed at which it has expanded to
that size.

Direct measurements of the advance speeds inferred from the proper
motion of the hotspots in compact symmetric objects (CSOs) have been made
by \citet{Ows98a} and \citet{Ows98b} which find them to be rapid (0.2\,$c$
-- 0.25\,$c$).  The model for source expansion of \citet{Fal91}, with the
sources expanding into poor-group type environments [e.g.\ as
parameterised in \citet{Blu99a,Blu99b}], predict such speeds for the
earliest stages of a source's life.  This model predicts that these speeds
reduce gradually, by an order of magnitude, as the sources age.  We now
consider speed estimates for these older classical double radio sources
which are the focus of this paper.  Many of these speed estimates have
come from consideration of side-to-side asymmetries which are presumed to
arise from light-travel-time effects.  These effects arise because the
lobe nearer to the observer is seen at a more recent epoch than the other
lobe so it is seen when it is older and hence, in the absence of other
influences, longer.

\begin{figure}[!h]
\begin{picture}(50,380)(0,0)
\put(-50,-35){\includegraphics{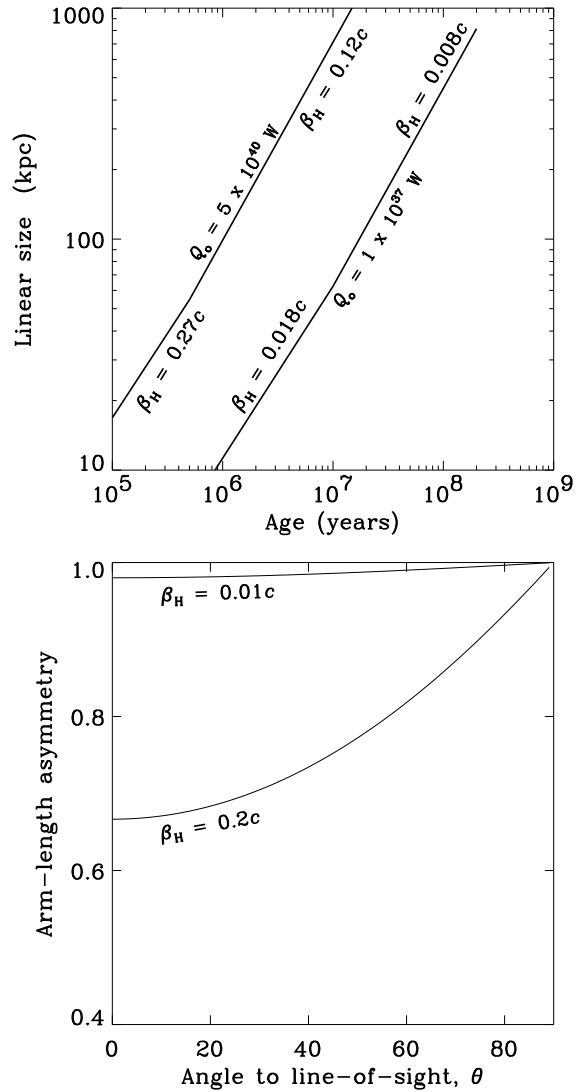}}
\end{picture}
{\caption[junk]{\label{fig:dist_age} {\em Upper plot:} Linear size of
radio source (in kpc) versus source age (in years); the left curve is for
a source each of whose jets carry a bulk kinetic power $Q_{\rm o} = 5
\times 10^{40}\,{\rm W}$ and the right curve is for a source with $Q_{\rm
o} = 1 \times 10^{37}\,{\rm W}$, derived according to the model of
\citet{Fal91} and assuming the same environments as for
Figure~\ref{fig:tmin}.  The higher powered source reaches a size of 1\,Mpc
after only $\sim 1.5 \times 10^7$ years while the lower powered source
reaches this size after $\sim 3 \times 10^8$ years.  The faster expansion
speed of the higher powered source should manifest itself in more dramatic
side-to-side asymmetries due to light-travel time effects as shown in the
lower plot; {\em lower plot:} the predicted arm-length asymmetries
expected for two different example advance speeds.  }}
\end{figure}

The faster a source is expanding the more likely it is to exhibit
asymmetries in arm-length, as expressed in Equation~\ref{eq:asym} and as
shown in the lower plot of Fig.~\ref{fig:dist_age}.  The observed
asymmetry in arm-length (quantified as $x$) of an intrinisically
symmetrical radio source expanding from the centre of a spherically
symmetrical environment, each of whose lobe-lengths is growing at speed $c
\beta_{\rm H}$ and whose jet-axis is oriented at angle $\theta$ to the
line-of-sight is given by:
\begin{equation}
\label{eq:asym}
x = 
\displaystyle{
\frac{(1-\beta_{\rm H} \cos\theta)}{(1+\beta_{\rm H}\cos\theta)}
}
\end{equation}

\citet{Lon79} measured the arm-length asymmetries of classical doubles
from the 3C catalogue and from the maximal asymmetries found, deduced that
there was an upper limit of 0.2\,$c$ to the advance speeds of these
objects.  A similar and subsequent analysis by \citet{Ban80} found that
half the classical double sources had advance speeds between 0.1\,$c$ and
0.4\,$c$.  More recently, \citet{Bes95} performed a similar analysis on
objects from 3C but when they allowed for the misalignment of up to
10\,deg out of collinearity for the two arms of a radio source found that
the distribution of arm-length asymmetries was best reproduced by a {\em
mean} advance speed of 0.2\,$c$, with some advance speeds extending up
to 0.4\,$c$.  A more realistic analysis was performed by \citet{Sch95} who
avoided attributing asymmetric arm-lengths to light-travel-time effects
unless there was independent evidence of relativistic effects from the
detection of a jet.  Mindful that environmental asymmetries have been
shown to cause significant asymmetries in
arm-lengths\footnote{\citet{McC91} found that the side of a source with a
shorter lobe was invariably found to be that with brighter emission-line
gas.} we conclude that arm-length asymmetries which are caused by effects
other than light-travel time effects must {\em not} be included in this
type of analysis.  Therefore we adopt Scheuer's upper limit that the
maximum advance speed of (fairly small and powerful) classical doubles is
likely to be $0.03c$.  Such a speed, when taken together with the physical
sizes of classical doubles (e.g.\ several 100s of kpc) such as those
studied by Alexander and Leahy (1987) give ages of $>\,10^8$ years.  These
ages are an order of magnitude larger than the ages Alexander \& Leahy
inferred by equating the spectral ages of the strips of lobe closest to
the core with the age of the radio source.

The largest radio sources (the so-called giants which have projected
linear size $D > 1\,{\rm Mpc}$) show no indication of being larger because
they have expanded more quickly\footnote{Although examples of a distinct
situation, rapidly expanding GHz-Peaked Spectrum (GPS) sources associated
with Mpc lobes, are known e.g.\ \citet{Sch99}.} than shorter radio
sources: \citet{Ish99} found for their sample of giant radio sources that
these are not significantly {\em more} asymmetric than sources with
smaller physical sizes.  If the giant radio sources do expand at speeds no
faster than the value quoted above, then their large linear sizes must be
achieved by higher ages, i.e.\ \gtsim\ a few $10^8$\,years, potentially
exacerbating the problem of source ages being highly in excess of the
maximum radiative lifetime for synchrotron-radiating particles.

The spectral ages of giants studied by \citet{Mac98} and \citet{Lac93}
(all with radio luminosities $< 10^{26} {\rm W\,Hz\,^{-1}\,sr^{-1}}$ and
likely to have jet-powers $\sim 10^{37}\,{\rm W}$) were found to be 3 -- 4
$\times 10^7$\,yr which if identified with their source ages would imply
speeds of \gtsim\,$0.1\,c$.  Given the maximum speed-limit identified by
\citet{Sch95} for the most powerful classical doubles, taken at face value
this would appear to go against the intuitively compelling idea
(substantiated by the dimensional analysis of \citet{Fal91}: see
Fig.~\ref{fig:dist_age}) that less powerful sources grow less quickly than
more powerful sources.

\section{How far away from the hotspot can the lobe material radiate?}
\label{sec:taildock}

\citet{Jen76} pointed out that if synchrotron cooling played a part in the
spectral shape of extended lobes, lobes should be observed to extend
further at low-frequency than at high-frequency.  This rarely seems to be
the case \citep{Blu99c}.  Rather, the extent of the lobe seems to depend
on the {\sl luminosity} of the radio source: low-$P$, large-$D$ sources
[like those studied by \citet{Ale87b}] invariably seem to have their lobe
emission extending all the way back to the core.  Evidence that this
strong structural dependence on luminosity was the case was presented by
\citet{Jen77} who identified that the more luminous classical doubles had
a higher fraction of flux-density originating in the outermost regions
(within 15\,kpc, i.e.\ the head and hotspot regions) than lower luminosity
sources.  This is very pronounced in high redshift radio galaxies where
the emission seen at GHz frequencies comes very largely from the hotspots
and very little from the lobes [see example maps of such high redshift
radio galaxies from the 6C* sample in \citet{Blu98a}].  In contrast, those
high-$P$, small-$D$ sources [such as those studied by \cite{Liu92}] seem
to show lobes extending only say, half-way back from the head to the core.
(Although sometimes in these powerful sources with docked tails, faint
emission can be still be seen close to the core \citep{Lea89}.  However,
while Leahy et al found this emission to be very {\em faint}, it was not
found to be {\em steep}.  Rather it was characterised by emission with a
similar spectral index to that measured somewhat closer to the hotspot.)

Thus it is in the {\em low}-luminosity sources that the confrontation of
radiative lifetime and source age is most problematic: GHz lobe emission
in these objects, including at points furthest from the hotspot, is seen
from synchrotron-emitting particles with radiative lifetimes \ltsim\ 10\%
of the source age.


\section{How to re-interpret/reconcile spectral ages}
\label{sec:mix}

There are two main approaches to the re-interpretation of spectral ages to
reconcile them with source ages: i) lowering the magnetic field throughout
the lobe below the equipartition values traditionally used, so that the
ages inferred from the spectral ageing method are higher and more
consistent with their true ages (\S\ref{sec:sub_equip}) and ii) mixing
plasma of different ages and particularly mixing the plasma furthest from
the hotspot with recently accelerated particles (this approach takes a
number of different forms and we discuss six of them in
\S\ref{sec:backflow} --- \S\ref{sec:stream}).

\subsection{Sub-equipartition $B$-field in the lobe-heads}
\label{sec:sub_equip}
\citet{Ale96} outline a means by which the expansion speed implied by the
spectral ageing analysis of Cygnus A might be made more than an order of
magnitude lower in order to be consistent with its dynamically-inferred
expansion speed based on the assumption of ram-pressure confined hotspots.
If the magnetic field strength in the lobe were below the value which
would be in energy equipartition with the particle pressure then the
spectral ages\footnote{Though note this still ignores the consequences of
expansion losses.} would be higher and less discrepant with the dynamical
ages than if equipartition magnetic field strengths were used.  The
expansion speed $v_{\rm h}$ inferred from spectral ageing depends on the
assumed magnetic field strength $B$ as $v_{\rm h} \propto B^{3/2}$, and
thus to reduce $v_{\rm h}$ by a factor of \gtsim\ 10 requires $B$ to drop
by a factor of \gtsim\ 5.

\citeauthor{Ale96} cited strong observational evidence for equipartition
in the hotspots \citep*{Har94}\footnote{\citeauthor{Har94} detected X-ray
emission associated with the hotspots which, if caused by the synchrotron
self-Compton process, requires a magnetic field strength very close to the
equipartition value given certain assumptions, (specifically that there is
no dominant presence of protons).}.  They suggested that it is the process
by which the plasma leaves the hotspot which causes the magnetic field
strength to go out of equipartition with the energy in the particles.
They aimed to model the flow out of the hotspot by steady diverging
streamlines along which magnetic flux is conserved (i.e.\ the
flux-freezing condition of $B \propto 1/A$ holds, where $A$ is the
cross-sectional area of some fixed loop) and along which mass is conserved
(i.e.\ $\rho v A$ is constant, where $\rho$ and $v$ are the density and
velocity along a streamline at some point).  They obtain a relationship
for the ratio of the energy densities in the particles ($u_{\rm p}$) and
in the magnetic field ($u_{\rm B}$) in terms of the velocities along the
streamlines.  This ratio is derived using relationships for the pressure
$p$ and density $\rho$ in terms of these velocities from \citet[eqns
83.16]{Lan59}.  They obtain the following:
\begin{equation}
\label{eq:pa}
\frac{u_{\rm p}}{u_{\rm B}} \propto \frac{p}{B^2} \propto
\frac{p}{\rho^2v^2} \propto 
\frac{1}{v^2/c_*^2}\frac{(1 - v^2/7c_*^2)^4}{(1 - v^2/7c_*^2)^6}
\end{equation}
where $c_*$ is the velocity at the point in the flow where $\rho
v$ is a maximum.  They approximate the velocity out of the
hotspot to be $c_*$.  \citeauthor{Ale96} claim that a drop in $v$
by a factor of $1/\sqrt{2}$ readily gives the required increase
in $p/B^2$ of a factor of 5.  In fact their formula (shown here
in equation~\ref{eq:pa}) shows that to achieve this increase in
$p/B^2$ requires the bulk flow velocity to decrease from its
initial value by a factor of 0.4.  

\subsection{Fast bulk backflow}
\label{sec:backflow}

In two spectral ageing studies \citep{Ale87b,Liu92} the authors found that
the spectral ages they derived implied advance speeds which were
unacceptably high, even not accounting for adiabatic losses, in the light
of the then upper-limit\footnote{See dicussion of more up-to-date and
tighter constraints in \S\ref{sec:sourceage}.} to head advance speeds
of \citet{Lon79} of $\sim 0.2\,c$.  As pointed out by \citet{Win80} the
speed at which a given strip of plasma in the lobe separates from the
hotspot is the anti-vector sum of the oppositely directed head advance
speed and the backflow speed in the frame of the host galaxy.  Bulk
backflow speeds of $\sim 0.1c$, i.e.\ comparable with the advance speeds,
were invoked to alleviate the requirement for such high advance speeds.

\citet*{Lea89} find that the central regions of lobes have much less
distortion in the higher-powered sources than do those of lower-powered
sources.  They suggest that this means that there is {\em less} backflow
in more powerful sources.  If the spectral age estimates were true this
would be in contradiction with \citet{Ale87b} and \citet{Liu92} who found
that the lobe speed $v_{\rm l}$ (sum of advance and backflow speeds
deduced from spectral ageing) increases with radio luminosity $P$, as
$v_{\rm l} \propto P^{0.33}$ assuming that the advance speeds are as
deduced by \citet{Sch95} (see \S\ref{sec:sourceage}).

A high bulk backflow speed requires some physical sink for the backflowing
material and a few examples of `X'-shaped sources are known where this may
be the case \citep{Lea97,Den99}.  However, there appears to be no evidence
of powerful classical double radio sources showing {\em any} excess lobe
structure indicative of backflow at low-frequencies like 74\,MHz than at
GHz frequencies \citep{Kas93,Blu99c}.  There are no examples known of lobe
emission near the core betraying signs of compression (for example by
enhanced surface brightness compared to the rest of the lobe).

In the context of a leaking hotspot model and the complex magnetic field
structure in the vicinity of the hotspot it seems unlikely that as plasma
emerges from the hotspot in the frame of the host galaxy it has a huge
{\sl bulk} flow away from the hotspot towards the core.

Although some simulations have shown rapid bulk backflow speeds \citep[see
e.g.\, ][]{Nor82}, these speeds are strongly influenced by the boundary
conditions used: open boundary conditions in the plane from which the jet
emanates, and through which lobe material may exit, will enforce rapid
flow away from the hotspot.  Simulations with more appropriate boundary
conditions can still show fast, but filamentary, backflow in the very
early stages of a source's life.  State-of-the-art simulations have
much to offer in this area.

\subsection{Turbulent (convective) backflow}
\label{sec:turble}

A variant on the bulk backflow model \citep{Lea89} is to
assume that the turbulent mixing within the lobe of
freshly-injected plasma and older plasma is sufficiently
large-scale and fast that efficacious transport of highly
energetic particles to great distances away from the hotspot is
achieved.  In practice, this requires the effective velocity of
transport to still be of the same order as the invoked bulk
backflow speeds discussed in \S\ref{sec:backflow}.

\subsection{Particle re-acceleration}
\label{sec:reaccel}

An alternative way of introducing highly energetic particles into
regions of the lobe close to the core of a radio source is to
invoke that the hotspot is not the only source of high-$\gamma$
particles, but that within the lobe itself re-acceleration of
particles occurs.

This has been invoked by \citet{Par99}, in the context of FR\,I
\citep{Fan74} radio sources, as a way to explain the low values of
spectral ages in comparison with the dynamical ages.  It has also been
invoked by \citet{Car91} in the case of Cygnus A to explain spectral
shapes observed near the hotspot which do not exhibit symptoms of
straightforward synchrotron losses.

\citet{Eil89} included {\em in situ} electron re-acceleration both by
Alfv\'{e}n waves and by a Fermi mechanism in their model radio sources
which were evolving in a uniform medium.  They found this gave little
global influence on the luminosity behaviour.  This may be because in
practice it is hard to achieve re-acceleration mechanisms in the lobe
whose efficacy is comparable with that of the hotspot, and the rate at
which it is able to supply energized particles.  It is even harder if one
considers that the particle acceleration mechanism has to generate
higher-$\gamma$ particles than are needed to radiate in the higher
magnetic field region of the lobe-head, a point to which we return in
\S\ref{sec:specgrad}.  Independent evidence that particle re-acceleration
is {\em not} occuring has been found by \citet{Rud94} and \citet{Kat94}
and is discussed in \S\ref{sec:katz}.

\subsection{Freezers}
\label{sec:freezer}

Eilek, Melrose \& Walker (1997) have considered possible physical
mechanisms which would cause radio sources to appear younger than their
implied dynamical ages.  They invoked that if a relativistic electron
spends most of its time in a low-field region then diffuses into a
high-field region in which it must radiate more efficiently than in a
low-field region, on average it loses energy more slowly than if it were
continuously in the high magnetic field region\footnote{Such an idea of a
`magnetic freezer' was first invoked by \citet{Sch89} in the context of
low magnetic fields in jets to preserve an extremely high Lorentz factor
population of particles from rapid synchrotron losses.}.  If there are
substantial inhomogeneities in the $B$-field and a sufficient quantity of
energetic particles were somehow contained within a low magnetic field
state then this could in principle occur.

The filamentary nature of radio lobes seen in the best quality total
intensity images \citep[e.g.\ Cygnus A by][]{Car91} is circumstantial
evidence that there are discrete regions of high and low magnetic fields
within lobes.  However, we note that it is only with the certainty of
containment of the fast synchrotron particles that this could be a viable
mechanism (see \S\ref{sec:stream}).

\subsection{Low pitch-angle reservoir}
\label{sec:pitch}

\citet{Spa79} pointed out that electrons with small pitch-angles do not
suffer significant synchrotron losses but would diffuse in pitch-angle due
to the influence of irregularities in the magnetic field he envisaged
being superimposed on a uniform field longitudinal with the lobe.  He
suggested that the loss via synchrotron cooling of energetic electrons in
high pitch-angle states is balanced by re-supply from the low pitch-angle
reservoir.  However, it is unlikely that the relative fraction of low
pitch-angle particles is sufficient to replenish the rather larger
fraction of rapidly radiating high pitch-angle electrons, as the fraction
of particles having pitch-angle $< \theta$, for an isotropic distribution,
is given by $1 - \cos\theta$.  Moreover, such a model relies on having an
essentially longitudinal magnetic field (for which there is no
observational evidence) and which would be difficult to sustain on
theoretical grounds (anchoring such magnetic flux would be very
difficult).

\subsection{Fast streaming}
\label{sec:stream}

We now consider whether relativistic particles can stream through
low-energy plasma to penetrate those parts of the lobe whose original
supply of energetic particles have long since radiated their energy away.

Streaming of high-$\gamma$ particles parallel to field lines is of course
extremely fast and can proceed at speeds close to $c$.  In contrast motion
perpendicular to the local field lines will be extremely slow.  Since
there is no observational basis for magnetic fields purely longitudinal
with the lobe (see \S\ref{sec:pitch}) and tangled fields (albeit on an
unknown variety of scale-lengths) are more likely, it is necessary to
consider mechanisms for the scattering of particles to randomize their
directions so that streaming parallel to the local field configuration may
occur.

The anomalous diffusion mechanism of \citet{Rec78} involves the random
walk of the particles taking place across randomly oriented regions of
magnetic field.  Following \cite{Duf95} and assuming a tangling scale of
10\,kpc, in $10^6$\,years the r.m.s.\ distance diffused is $\sim$ a few
100\,kpc.  This distance is comparable with the arm-lengths of the
large-$D$, low-$P$ sources in the spectral ageing study of \citet{Ale87b}
and the timescale is less than the radiative lifetime of synchrotron
emitting particles in these sources (\S\ref{sec:radlife}).

Such scattering mechanisms are necessary to explain the difficulties of
plasma containment seen in laboratory plasmas \citep{Den93} and there is
no basis for believing that astrophysical plasmas should not have the
tangled configurations which cause the particles to be scattered.

\citet{Win80} claimed that there is an upper limit to the mixing speed of
$4 \times$ the advance speed based on the assertion that otherwise a
uniform population of electrons would be seen in all parts of the lobe;
such an analysis neglects the influence of the magnetic field
configuration (specifically the gradient in strength from hotspot to core)
within the lobe, discussed in \S\ref{sec:specgrad}.  The speeds which we
discuss here should be regarded as penetration speeds rather than mixing
speeds.

\section{Energy transport and the rarity of dead radio galaxies}
\label{sec:relic}
It is clear that there are some serious drawbacks with the assumptions of
a significantly sub-equipartition $B$-field (\S\ref{sec:sub_equip}, see
also \S{sec:equip}) {\em within the lobe-heads}, fast bulk backflow
(\S\ref{sec:backflow}), convective backflow (\S\ref{sec:turble}) and a low
pitch-angle reservoir (\S\ref{sec:pitch}).

However, the scenarios for replenishing the lobes with energetic particles
long after the hotspot has passed by outlined in \S\ref{sec:reaccel}
(particle re-acceleration), \S\ref{sec:freezer} (freezers) and
\S\ref{sec:stream} (fast streaming), in principle can resolve any
discrepancies implied by spectral ageing.  However, whatever the details
of the physical mechanisms by which the scenarios discussed in
\S\ref{sec:reaccel} (particle re-acceleration) and \S\ref{sec:freezer}
(freezers) would be governed, it is not clear that either of these
processes would know when to stop.  In other words, there is potentially a
problem with a surfeit of sources having highly radiant lobes long ($>
10^8$\,years) after jets and hotspots have switched off.

Examples of relic classical doubles are rare, with only a very few
examples known \citep{Cor87,Har95,Har93}.  \citet{Ril89} has used the
sensitivity of the 6C survey to investigate whether any extended sources
were unwittingly missed by the revised 3CR survey of \citet{Lai83}.  Apart
from a few slight corrections due to measurement errors and confusion, no
such sources were found.  When Riley lowered the flux-limit from 12\,Jy at
151\,MHz to 9.5\,Jy, while the sample size increased by 50\%, all of these
sources were found to be bona fide classical doubles.  \citet{Gio88} find
that there are just 3 -- 4 \% of their sample of radio sources selected
from the B2 and 3C catalogues which are relics.

In the example relic classical double studied by \citet{Cor87}, IC\,2476,
lobes of smooth extended low-surface brightness emission are seen in
images with greater than half-arcmin resolution, which straddle a $z =
0.027$ galaxy.  Higher resolution images resolve out these lobes
completely, yet find no evidence of hotspots or a core or jet.  It is very
likely that at non-zero redshifts the absence of relic radio galaxies has
its explanation in the {\em `Youth-Redshift degeneracy'} we described in
\citet{Blu99b}.  This is the effect in which a combination of declining
luminosity-with-age for radio galaxies\footnote{These arise in all
reasonable models of radio galaxy evolution and environment, even while
the jet continues to deliver a {\em constant} bulk kinetic power.}  and
the application of a survey flux-limit mean high-redshift radio galaxies
are not seen at late stages in their lives.  However, this explanation
does not help to explain the deficit of relic radio galaxies in the local
($z < 0.5$) Universe, where the effect of the flux-limit is not so
drastic.

The expansion of the low-$\gamma$ plasma lobe can only occur on timescales
consistent with the sound speed so this cannot be a rapid process.  A
rapid decay time does however come from considering the brevity of the
radiative lifetimes of the synchrotron emitting particles.

We suggest that the role of fast streaming via anomalous diffusion of
energetic particles throughout radio lobes might make a new contribution
to the explanation for the rarity of `dead' radio sources found to date in
surveys of the local Universe.  If there were a sudden cessation in the
supply of bulk kinetic jet-power, a radio source might begin to fade from
view within a time-scale comparable with the maximum radiative life-times
of the synchrotron particles (\gtsim\ $10^6$ years) of the hotspot ceasing
to inject high-$\gamma$ particles into the lobe.  Though the magnetic
field can still be present as long as the low-$\gamma$ plasma lobe is
confined, high-$\gamma$ particles are no longer injected into the lobe to
replenish the synchrotron-radiating population.

\section{A refined spatially-resolved model of radio sources}
\label{sec:refine}

Upto now it has always been assumed that the high-$\gamma$ particles are a
smooth continuation of the ones which govern the magnetic field and into
which the magnetic field is frozen, and that the localisation of the
high-$\gamma$ particles is co-spatial with the magnetic field.  This has
yet to be justified in the context of astrophysical plasmas.

\subsection{Assumptions about hotspot output and the nature of
the lobe}
\label{sec:refinehotspot}

We assume that the bulk backflow velocity out of the hotspot is
very close to zero in the frame of the host galaxy.  We picture
the hotspot dumping successive elements of plasma.  These
elements relax into the lower pressure of the head region and,
bringing a new supply of particles, the volume of the head region
increases and the source expands.

In the radio source model we developed in \citet{Blu99a}, each element of
hotspot material which transfers to the lobe-heads has an energy
distribution which is characterized by two adjoining power-laws, the
low-energy regime having a frequency spectral index\footnote{We use the
convention for spectral index ($\alpha$) that $S_{\nu} \propto
\nu^{-\alpha}$, where $S_{\nu}$ is the flux density at frequency $\nu$.}
of $\alpha = 0.5$ and above the break frequency $\nu_{\rm B}$ the
frequency spectral index is $\alpha = 1$.  Where the break frequency
occurs depends on the dwell-time for which particles have resided in the
enhanced magnetic field which has built up in the hotspot.  Thus with
elements of the hotspot escaping into the lobe which have had different
dwell-times in the hotspot, a summation of spectra are injected into the
lobe which together comprise a {\em curved injection spectrum}.  Note that
such a model resolves the `injection index discrepancy' discussed by
\citet[ \S5.2.2.2]{Car91}.  \citeauthor{Car91} said that the low-frequency
spectral index measured in the lobe should reflect {\em either} the
$\alpha = 0.5$ spectral index of the hotspot (if expansion losses were
small) {\em or} the $\alpha \sim 1$ spectral index of the hotspot seen at
higher frequencies (if the expansion losses were large).  In fact they
measure a low-frequency spectral index of 0.7.  We remark that
the low-frequency spectral index should thus not be identified with the
injection spectrum; this might lead one to spuriously conclude that the
injection spectrum varies along the lobe and hence throughout the lifetime
of a source.

These elements dumped into the lobe-head we suggest may be
identified with the turbulent eddies in the hotspot model of
\citet{DeY99} which have escaped out of the hotspot.  As long as
they are on sufficiently small spatial scales these eddies will
have magnetic field strengths in equipartition with their
particle energies \citep{DeY99}.  Equipartition of the magnetic
energy density and the particle energy density is maintained in
the hotspot to head transition as long as the mixing ---
energized by the eddies from the hotspot --- persists.


We assume that once an element of plasma is `dumped' by the hotspot it
will gradually expand to the ambient pressure, that of the head.  We
regard this input of particles as constituting a new slab of the lobe
whose addition lengthens the lobe.  Subsequent injections of plasma into
the lobe add further slabs each of whose pressure is similar to that of
the slabs either side.  Subsequent expansion of these volumes is entirely
transverse since they are approximately pressure matched front and back.
If the pre-expansion magnetic field is purely longitudinal then the
magnetic field strength in these slabs will follow $B \propto 1/A$ and if
it is tangled then it is approximated by $B \propto 1/\sqrt{A}$ where $A$
is the cross-sectional area of the slab.  This causes a decline in the
magnetic field strength, a point to which we return in \S\ref{sec:equip}.

The collection of many slabs constitutes a body of plasma which purveys
the magnetic field.  We posit that the slabs of plasma are dumped by the
hotspot, and once the radiative lifetime of their particles has expired,
will make no further direct contribution to the synchrotron radiation.
However, their contribution is as a matrix of low-$\gamma$ material
supplying a magnetic field which subsequently injected high-$\gamma$
particles penetrate (perhaps via the \citet{Rec78} mechanism, see
\S\ref{sec:stream}) and radiate in.

\subsection{How the lobe magnetic field evolves}
\label{sec:equip}

\begin{figure}[!h]
\begin{picture}(50,230)(0,0)
\put(-14,-120){\includegraphics{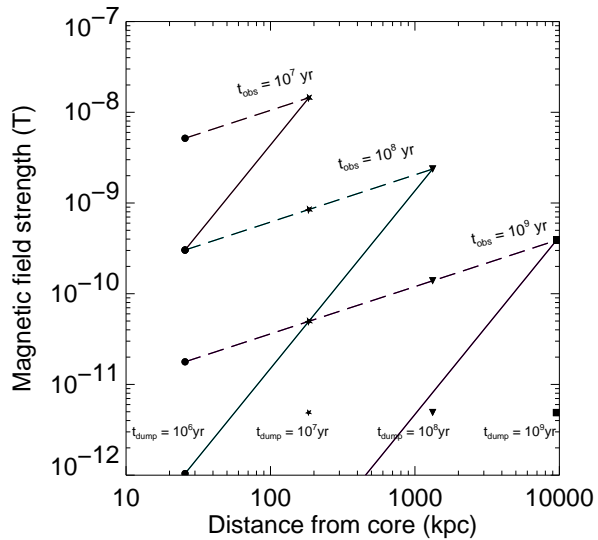}}
\end{picture}
{\caption[junk]{\label{fig:bfield} Magnetic field strength in a radio lobe
at a given distance away from the core.  The solid sloping lines are
calculated for the model described in \S\ref{sec:equip}, where
flux-freezing is invoked as the lobe expands transversely.  The dashed
sloping lines show the magnetic field strength throughout the lobe if the
field configuration is such that $B \propto 1/\sqrt{A}$.  The upper set of
lines show the various magnetic field strengths if the source is observed
aged $10^7$\,years, the middle set if the source is aged $10^8$\,years,
and the lower set if the source is aged $10^9$\,years.  As the source ages
the $B$-field falls.  Each set of lines corresponds to a source with each
jet carrying a bulk kinetic jet-power $Q_{\rm o} = 5 \times 10^{39}\,{\rm
W}$.  The same environmental parameters are used as for
Figure~\ref{fig:tmin}. }}
\end{figure}

We have used our non-self-similar model for radio galaxy evolution
(developed in \citet{Blu99a}) and refined it as in
\S\ref{sec:refinehotspot}.  We emphasize that in this model equipartition
is preserved in the transfer of plasma from the hotspot to the head.  At
successive points throughout a source's lifetime we consider the volume of
plasma injected into the lobe from the hotspot in the preceding $10^3$
years.  The length of this slab of plasma is taken as the distance
advanced by the head of the source during that $10^3$ years.  We imagine
that a slab of plasma, being surrounded on either side by slabs of plasma
at very similar pressure, can expand only transversely into the IGM.  Note
that this is the same geometrical configuration as envisaged by
\citet{Chy97}.  Transverse expansion is what governs the strength of the
tangled magnetic field.  With an assumed pressure gradient away from the
head which falls off with the square of the distance (thus giving a ratio
of pressure in the head to that half-way along the lobe of $\sim 4$ in
agreement with \citet{Kai99}), this simple model can give a drop in
magnetic field of over an order of magnitude in Mpc-scale radio sources
(see Fig~\ref{fig:bfield}).

We note that the $B$-field strength in almost all of the most radiant
regions of the lobe and head is nothing like the factor of 5 below the
equipartition value at the head required by \citet{Ale96} to explain the
spectral age discrepancy in Cygnus A.  The region of the lobe nearest the
core has the lowest $B$-field at $t_{\rm obs}$; however, when the source
was aged \gtsim\ a few $10^7$ years the $B$-field of this plasma was
$\sim$ an order of magnitude higher (Fig.\ \ref{fig:bfield}) synchrotron
cooling would have severely depleted the high-$\gamma$ particles injected
simultaneously.  Very low $B$-fields in the lobes near the core at $t_{\rm
obs}$ cannot reconcile spectral and dynamical ages.

\section{Spectral gradients}
\label{sec:specgrad}

Spectral gradients along the ridgelines of classical doubles, in
the sense that spectra are steeper nearer the core than at the
head of the lobe, are a common, though not ubiquitous, feature of
these objects. An early example of a spectral gradient was found
by \citet{Bur77} in 3C452 where the spectral index between
1.4\,GHz and 5.0\,GHz was found to steepen from 0.7 near the head
of the source to 1.7 near the centre of the source.  Burch
attributed this behaviour to increasing synchrotron losses with
increasing distance from the core.  The spectral steepening in
the sources studied by \citet{Ale87b} at GHz frequencies was
directly interpreted as increasing synchrotron losses with
increasing time elapsed since injection from the hotspot.  This
assumption paved the way for their break frequency fitting and
spectral ageing analysis.

Cygnus A itself has been shown by \citet{Ale84} to have lobe
spectra which steepen away from the hotspots between frequencies
of 1\,GHz and 5\,GHz, with the spectral index reaching a maximum
of about two in the most extended structure. They interpreted
these variations as the results of synchrotron ageing, making it
possible to estimate the speed of advance of the hotspots.
However, this same source has a spectral index gradient between
74\,MHz and 327\,MHz \citep{Kas96} of $\Delta\alpha$ \gtsim\ 0.3.
It is hard to see that this low-frequency gradient could have the
same origin as that which \citet{Ale84} suggested for the GHz
regime.

Spectral gradients are seen to extend to sub-GHz frequencies in other
examples as well \citep[e.g.][between 610\,MHz and 1.4 GHz]{Jag87}.

If particles can diffuse distances comparable with lobe-lengths in a few
$10^6$ years, then the constant replenishment of energetic particles
suggests it is unlikely that synchrotron ageing is the sole cause of the
observed spectral steepening.  Undoubtedly this plays a part in some
short, very powerful sources however.

\citet{Wii90} have questioned both the assumption that the magnetic field
is uniform throughout the lobe and the traditional useage of an estimate
of the magnetic field strength in spectral ageing analyses made in a
region near the hotspot\footnote{The motivation for this previous useage
is that this is a region in which the low-frequency spectral index should
most reflect the injection index, but see the caveat about this in
\S\ref{sec:refine}.}. They suggest that a lower $B$-field should be found
closer to the core, rather than the higher `pre-expansion' $B$-field found
nearer the hotspot.  \citeauthor{Wii90} say that a lower magnetic field
will result in a higher break-frequency, and use this fact to explain the
`anomalous' spectral gradient in 3C234 found by \citet{Ale87a}.  In this
particular case, using their preferred $B$-field estimates gives spectral
ages which are higher by a factor of only 1.5 than those of
\citet{Ale87a}.

We now consider the consequences on the spectral gradient of a magnetic
field which declines along a lobe from the head towards the core, given
the drop by $\sim$ an order of magnitude shown in Fig.~\ref{fig:bfield}.
For a given observing frequency, say 8\,GHz, the emission arising from the
higher magnetic fields near the head of the source will arise from {\em
lower} Lorentz factor particles than the emission coming from lower
magnetic fields closer to the core which inevitably require {\em higher}
Lorentz factor particles for radiation at the same frequency.  If the
underlying energy distribution of synchrotron particles is curved (an
inevitable consequence of the leaky hotspot model described in
\S\ref{sec:refinehotspot}) then the regions near the cores of these objects
will show steeper spectra than those near the head.

\begin{figure}[!h]
\begin{picture}(50,220)(0,0)
\put(-150,-315){\includegraphics{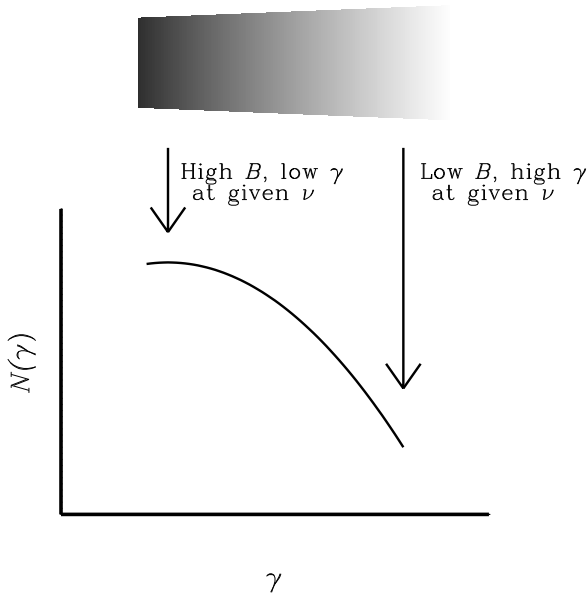}}
\end{picture}
{\caption[junk]{\label{fig:fish} Schematic illustration of the
gently decreasing gradient in magnetic field strength along the
lobe away from the head region, and how for an observation at say
8\,GHz, increasingly high Lorentz factor particles will be
sampled in regions with lower magnetic field.  }}
\end{figure}

Note that a spectral gradient will only be seen with a combination of a
magnetic field gradient and a curved spectrum.  There do exist sources
with no spectral index gradient \citep[see][]{Jag87}.  It may be that they
have a magnetic field gradient, but the spectrum of particles supplied by
the hotspot is (e.g.\ because the hotspot has such a high $B$-field) only
described by $\alpha = 1$, i.e.\ any break in the hotspot spectrum is not
seen by an amount $\delta \alpha = 2 a_{2} \log_{10}(B_{\rm head}/B_{\rm
lobe})$ where $a_{2}$ is the coefficient of the curvature term in the
frequency spectrum (c.f.\ Blundell et al 1999a) --- see schematic
illustration in Fig.\ \ref{fig:fish}.  

\citet{Car91} find that the bright filaments of Cygnus A have {\em higher}
break frequencies than their environs and throughout the source they find
a correlation of total intensity and break frequency.  We interpret this
as confirmation that at a chosen observing frequency the higher $B$-field
regions in a lobe will be `illuminating' a lower energy regime of the
underlying particle distribution, which will have a flatter spectrum
(hence higher `break') than the higher energy regimes illuminated by lower
$B$-fields (see Figure~\ref{fig:fish}).  This is strong evidence in favour
of replenishment rather than static spectral ageing because it is the
opposite effect to that predicted by synchrotron ageing.

\section{Independent evidence for quantitative magnetic field
gradients} 
\label{sec:newbit}

We now consider quantitative estimates from the literature of magnetic
field strengths and their comparisons with equipartition inferred values.

\citet{Mur99} have found for the very young (age \ltsim\ $10^5$ years)
sources they studied that the values of equipartition magnetic field
strength which they derive are consistent with the magnetic field
strengths responsible for the low-frequency turnovers via the synchrotron
self-absorption mechanism.

\citet{Har94} show that the magnetic field required to give synchrotron
emission observed in the hotspots (where the number of particles has been
normalised by interpreting the X-ray emission from the hotspots as
synchrotron self-Compton, i.e.\ it is the photons emitted by the
synchrotron electrons which are up-scattered) is very close to the
equipartition value inferred in the usual way.

Further removed from the hotspots, X-ray emission at $\sim 2$ keV has been
detected associated with radio lobes which is likely to be CMB photons
up-scattered by synchrotron particles with Lorentz factors of $\sim 1000$,
as follows:

\begin{enumerate}
\item \citet{Fei95} and \citet{Kan95} independently find for Fornax A that
the ratio of the X-ray luminosity from the inverse Compton up-scattered
CMB photons taken together with the synchrotron emission from the lobes
imply a magnetic field strength very close to that which is inferred from
the usual equipartition arguments.

\item For 3C\,326, Tsakiris et al (1996) have inferred magnetic fields
which are the same as, or lower by a factor of 2 than, the magnetic field
strength derived from the usual equipartition calculation on the basis of
X-ray emission detected near the heads.

\item Subsequently for Cen B, \citet{Tas98} found that the magnetic field
strength implied by interpreting the X-ray emission as up-scattered CMB
photons gives a magnetic energy four times lower than the particle energy.
Such a wide discrepancy is most pronounced in the furthest regions from
the hotspots, while at the hotspots themselves, equipartition is in fact
preserved.
\end{enumerate}

\subsection{Pressure and magnetic field constraints near the core}
\label{sec:budget}

We now turn our attention to independent calibration of the pressure in
the lobes.  \citet{Lea99} have found that there is a considerable
discrepancy with the value of the lobe pressure inferred from the usual
equipartition argument and that obtained by fitting a King model to the
profile of the cluster X-ray emission surrounding Her A, assuming a
temperature and deducing the pressure which is confining the lobes.  This
is in the sense that the lobe pressure appears to be an order of magnitude
too low for pressure balance with the external confining medium.
\citet{Lea99} conclude that some non-radiating, low-energy contribution
(such as protons or very low-$\gamma$ particles) must be responsible for
the extra pressure.  There is considerable independent evidence that very
low-$\gamma$ particles must be present in the lobes: i) they are present
in the jets which feed the lobes [as required by the details of the
circular polarisation detections of \citet{War98}] ii) they are an
inevitable consequence of the synchrotron decay process.

If it is the case that the particles which emit the observed synchrotron
radiation represent a high-$\gamma$ tail which has penetrated through the
low-$\gamma$ matrix then it is clear that the standard minimum energy
calculation based on this emission, including that near the core, has a
very unsound basis.

Moreover, if it is the case that the particles which emit the observed
synchrotron radiation represent a high-$\gamma$ tail which has penetrated
through the low-$\gamma$ matrix then it is clear that the standard minimum
energy calculation based on this emission, including that near the core,
has a very unsound basis.  

For a lobe with a particle reservoir whose energy distribution is given
by
\begin{equation}
N(\gamma){\rm d}\gamma = K\gamma^{-(2\alpha + 1)}{\rm d}\gamma
\end{equation}
between all particle energies from $\gamma_{\rm min}$ to $\gamma_{\rm
max}$ residing in mean magnetic field strength $\overline{B}$ over volume
$V$ (assuming a filling factor of unity and no significant contribution
from protons) the total energy budget is given by:
\begin{equation}
\displaystyle
E_{\rm tot} = V K m_{\rm e} c^2 
              \ln \biggl[ \frac{\gamma_{\rm max}}{\gamma_{\min}} \biggr]
              + \frac{\overline{B}^2V}{2\mu_0}
\end{equation}
if $\alpha = 0.5$, and otherwise: 
\begin{equation}
\label{eq:totenergy}
\displaystyle
E_{\rm tot} 
= V K m_{\rm e} c^2 
  \biggl[ \frac{1}{\gamma_{\rm min}^{2\alpha - 1}} - 
          \frac{1}{\gamma_{\rm max}^{2\alpha - 1}} \biggr]
        + \frac{\overline{B}^2V}{2\mu_0}.
\end{equation}

For the case where $\alpha = 0.5$ the energy budget is governed by the
upper-limit to the energy distribution, but given that no known {\sl
lobes} have spectral indices as flat as this we do not consider this case
further.  For steeper spectral indices, Equation~\ref{eq:totenergy} shows
us that this depends essentially not at all on $\gamma_{\rm max}$
(hereafter we neglect this term) but is governed by $\gamma_{\rm min}$.  A
choice of $\gamma_{\rm min} = 50$ will under-estimate this contribution to
the energy budget by a factor of e.g.\ $50^{0.2}$ for $\alpha = 0.6$ if
the true value is $\gamma_{\rm min} = 1$.

On the naive assumption that at least for low energies the distribution
can be represented by a power-law, the other major unknown in this is the
normalisation of the number density of particles, $K$.  This has been
estimated in a couple of cases by inferring the number densities of the
low-$\gamma$ particles required to inverse Compton scatter photons from
the AGN radiation field upto the soft X-rays\footnote{We note in passing
that it is possible to misconstrue extended X-ray emission from powerful
high-$z$ classical doubles which has been upscattered from the powerful
QRF as arising from surrounding cluster emission.  Accurate morphological
information from high-resolution observations with Chandra will be a great
asset.}  in the cases of 3C\,219 \citep{Bru99} and Cen\,B \citep{Tas98}.
In the former case, \citet{Bru99} found that the number density of
particles must exceed the equipartition value by a factor of 10 (this is
derived assuming $\gamma_{\rm min} = 50$).  In the latter case,
\citet{Tas98} assume that $\gamma_{\rm min} = 1000$ which given the
evidence of inverse-Compton scattering off the AGN radiation field, denies
the presence of the $\gamma \sim 100$ particles responsible and is
probably not the best choice of $\gamma_{\rm min}$ given the evidence
discussed earlier in this section.  \citet{Bru99} recalulate the energy
budget for Tashiro et al's data using $\gamma_{\rm min} = 50$ and find
that a number density $\sim 50$ times higher than implied by the
equipartition value is required to explain their X-ray detections.

The observed synchrotron luminosity $P_{\nu}$, is given by:
\begin{equation}
P_{\nu} \propto V K \overline{B^{1 + \alpha}} \nu^{-\alpha}.
\end{equation}
The unknowns in this equation are famously $K$ (the normalisation of the
number density) and $\overline{B}$ (the mean magnetic field strength),
although the product $K \overline{B^{1 + \alpha}}$ is of course
determined.  If $K$ is found to be $\sim 10$ times higher than
equipartition values then the magnetic field strength giving rise to the
observed luminosity is lower by a factor of $\sim 10^{1/(1 + \alpha)}$
than the traditional equipartition calculation implies.  Note that the
constraints in number density from \citet{Bru99} and \citet{Tas98} pertain
to emission which is close to the core, i.e.\ in the oldest regions of the
lobes, precisely those regions which our simple model of \S\ref{sec:equip}
predicts to be those for which the synchrotron emitting population is most
poorly approximated as a smooth continuation from the low-$\gamma$
population.  

These results on the independent determination of magnetic field strengths
(for the hotspots, heads and lobes-near-the-core) are thus in accordance
with our simple model.   The fast-streaming picture of \S\ref{sec:stream}
means that the spectral shape of the high-$\gamma$ population can be
preserved throughout the lobe, but the normalisation (i.e.\ $K$) will
depend on the details of the diffusion mechanism.  At the heads of the
lobes the plasma is being viewed within a time-scale shorter than the
synchrotron radiative lifetime so the high-$\gamma$ population is a smooth
continuation of the low-$\gamma$ population.  However, near the core the
normalisation of the high-$\gamma$ population could substantially
underestimate the normalisation of the low-$\gamma$ population.  

There are some very important implications of our inference in this
section that the lobe pressures are roughly an order of magnitude larger
than those estimated by the standard minimum energy arguments.  We have
already considered this possibility in \citet{Wil99}: in the notation of
that paper, $f$ is the factor (greater than unity) which when multiplied
by the minimum energy value (calculated in the traditional way) gives the
total power which the jets have delivered since they first switched on
(see equation 4 in that paper) and has a value of $\sim 20$.  This value
is made up of several multiplying factors such as $f_{\rm min}$ (which
accounts for an excess energy in the particles compared with that in the
magnetic field), $f_{\rm geom}$ (which accounts for the unknown
deprojection of a radio-source onto the plane of the sky), $g_{\rm exp}$
(which allows for work done by the expansion of the radio source against
the external medium) and $g_{\rm ke}$ (which accounts for energy in the
bulk backflow of the lobe).  Our inference in this section is that the
bulk of the $f$ factor is contributed by $f_{\rm min} \sim 10$, leaving
room for only much smaller contributions from $g_{\rm exp}$ (probably
$\sim 2$), $f_{\rm geom}$ etc.  There is therefore no room for major
contributions to the pressure from protons, or for the filling factor to
be much lower than 1.  Furthermore, the greater total energy in the lobes
changes the normalisation of the inferred jet power.  Willott et al's
(1999) figure 7 shows that the effect of this change is to suggest that
the bulk power in the jets is then of the same order as the power radiated
away by the accreting central engine.  This rough equality of kinetic and
radiated outputs of an accreting black hole is a prediction of some models
\citep{Fal95}.

\section{Discussion of Katz-Stone \& Rudnick results}
\label{sec:katz}

Rudnick et al.\ (1994) and Katz-Stone \& Rudnick (1994) have made a new
type of analysis of the high-fidelity imaging data of Carilli et al.\
(1991) on Cygnus A.  They found evidence for a universal (curved) spectrum
which describes emission over the entire source.  That is, they take the
spectrum at various points throughout the source and then on the
$\log({\rm flux-density})$ {\em vs} $\log ({\rm frequency})$ plane they
perform translations to the different spectra which are parallel to either
one or other axis.  They found that these translations meant that all
these spectra would overlay on one another, i.e.\ they had precisely the
same shape.  This result has come from high-resolution images whereas
lower resolution images can give the impression of power-law spectra.  We
note that the most recent detections of the hotspots in Cygnus\,A in the
sub-millimetre, by \citet{Rob98} with SCUBA, imply a power-law spectrum
with spectral index $ \sim 1$ but with the comparatively low angular
resolution of SCUBA ($\sim 15^{\prime\prime}$, at 850 ${\mu}m$) it is
likely that these spectra will be flattened by the contribution of the
other hotspot components in the SCUBA beam.

Katz-Stone \& Rudnick found no evidence for the evolution of the electron
energy distribution function.  Neither did they find a distribution
resembling the \citet{Kar62} \& \citet{Pac70} model or the \citet{Jaf73}
model; nor do they find any evidence of the traditionally assumed
injection power-law distribution.

The translations of the spectra which Rudnick and Katz-Stone et al.\
performed are effectively moving around in age--$B$-field space (see
\citet{Kat94}).  The curve they find is telling us, albeit indirectly,
about the underlying particle distribution (which is what the hotspots
inject). The universality of the spectrum found for Cygnus A strongly
suggests that no particle acceleration processes have taken place at any
significant level since the shape of the original spectrum would not be
preserved by such a process.  Under these circumstances it is not a
surprise that the spectra should be the same in the lobes and the
hotspots.  This argues for a constant injection scenario [as in the
hotspot model of \citet{Blu99a}] throughout the fraction of a source's
life comparable with or greater than the radiative lifetimes of the
synchrotron-emitting particles.

The inclusion of fast-streaming of high-$\gamma$ particles via anomalous
diffusion helps to explain why \citet{Rud94} see the universal spectrum:
all parts of the lobe show us the energy distribution of particles as they
were when (or within a few $10^{6-7}$ years of being) first injected from
the hotspot.

A lobe cut into a sequence of strips will {\em not} just be telling us
about a sequence of differences arising just from the time of injection
(as is inherently assumed in the spectral ageing analyses) but mainly
about different and decreasing magnetic fields.  If fast streaming does
occur then this would be convolved with different Lorentz factors from
different times of injection.

\section{Concluding remarks}
\label{sec:conc}

The spectral ageing method can work well as long as the ages to be
measured are much shorter than the radiative lifetimes of the particles
which constitute the spectrum {\em and} when the history of the magnetic
field is well-approximated.  This has recently been demonstrated by
\citet{Ows98a}, \citet{Ows98b} and \citet{Mur99} for sources with ages
$10^2$ -- $10^3$ years and $10^5$ years respectively.  In the former
cases, direct verification of the advance speeds via proper motion
measurements of the hotspots with the VLBA is available and shows
impressive agreement with the spectral ages of a similar object derived by
Readhead et al (1996).  It is interesting in this regard that rather more
convincing broken spectra and spectral breaks are found in these cases
than for the spectra extracted from the much older classical doubles.

The possibility of replenishment of high-$\gamma$ particles via diffusive
streaming means that spectral ages in older and larger classical doubles
are even more meaningless than previous worries have suggested.

The best way of viewing radio lobe emission is via an evolving magnetic
field mediated by a low-$\gamma$ matrix which, according to its strength
and the observing frequency, will illuminate some small part of the
electron energy distribution as recently injected by the hotspot.  Such an
approach is supported by the observations of Carilli et al.\ (1991) that
the bright filaments of Cygnus A have flatter spectra than the surrounding
fainter emission.

\acknowledgments

K.M.B.\ thanks the Royal Commission for the Exhibition of 1851 for a
Research Fellowship.  We are are very grateful to Richard Dendy, Peter
Duffy, Torsten En{\ss}lin, Paddy Leahy and David De Young for stimulating
discussions and especially to Robert Laing and the referee Larry Rudnick
for a careful reading of the manuscript.

\end{document}